\documentclass[12pt]{article}
\usepackage{amssymb}

\input{tcilatex}

\begin{document}

\bigskip\ 

\bigskip\ 

\begin{center}
\textbf{ARE 1+1 AND 2+2 EXCEPTIONAL SIGNATURES?}

\textbf{\ }

\textbf{\ }

\smallskip\ 

J. A. Nieto\footnote{%
nieto@uas.uasnet.mx}

\smallskip

\textit{Facultad de Ciencias F\'{\i}sico-Matem\'{a}ticas de la Universidad
Aut\'{o}noma}

\textit{de Sinaloa, 80010, Culiac\'{a}n Sinaloa, M\'{e}xico}

\bigskip\ 

\bigskip\ 

\textbf{Abstract}
\end{center}

We prove that $1+1$ and $2+2$ target `spacetimes' of a $0$-brane are
exceptional signatures. Our proof is based on the requirement of $SL(2,R)$
and `Lorentz' symmetries of a first order lagrangian. Using a special kind
of $0$-brane called `quatl', we also show that the exceptional signatures $%
1+1$ and $2+2$ are closely related. Moreover, we argue that the $2+2$ target
`spacetime' can be understood either as $2+2$ worldvolume `spacetime' or as `%
$1+1$-matrix-brane'. The possibility that the exceptional $2+2-$signature
implies an exceptional chirotope is briefly outlined.

\bigskip\ 

\bigskip\ 

\bigskip\ 

\bigskip\ 

\bigskip\ 


Pacs numbers: 04.60.-m, 04.65.+e, 11.15.-q, 11.30.Ly

September 22, 2004

\newpage \bigskip

\textbf{1.- Introduction}

\smallskip\ 

In both mathematics and physics there are exceptional cases. The five
regular polyhedra, five exceptional groups, four division algebras, among
others, offer beautiful exceptions in mathematics [1]. On the other hand,
the four fundamental forces of nature and the five superstring theories [2]
provide with two examples of exceptional objects in physics. The dream in
this framework is to find a general theory of exceptions. For instance, $M$%
-theory [3]-[5] is one of the latest dreams in this quest in connection with
the five superstring theories.

One of the most important open problems in fundamental physics is to find
what the exceptional signatures of the space time are. Traditionally, in the
basic formulation of gauge field theories it is introduced by hand the
signature of $1-$timelike dimension and $3-$spacelike dimensions, that is $%
1+3$ signature. In Kaluza-Klein theory the situation is similar since one
chooses as starting point $T-$time, with $T=1,$ and $S-$spacelike
dimensions. In contrast supersymmetry offers an excellent mechanism to find
exceptional signatures. For instance, only with $T=1,S=9,$ $T=9,S=1$ or $%
T=5,S=5$ supersymmtry makes sense in superstring theory. Moreover,
supersymmetry application in $p$-brane theory leads to exceptions in both
time and space coordinates [6]. However these exceptions are too general in
the sense that are modulo some specific value. In this scenario it may seem
surprising to realize that one does not even need supersymmetry to find
exceptional signatures. Recent studies in two time physics [7] observed that
the symmetry $SL(2,R)$ in a first order classical lagrangian associated with
a $0$-brane is accomplished only if one considers a scenario with two times
and arbitrary spacelike coordinates. In this work, we analyze this
possibility more carefully founding that the $SL(2,R)$ symmetry, together
with a global Lorentz symmetry, leads to the interesting result that $1+1$
and $2+2$ dimensions are exceptional signatures.

At \ `human scale' our World appears to have $1+3$ signature. But why does
Nature make this signature an exceptional case? Recently some authors
[8]-[9] have proposed a possible explanation for this question. Their
proposals are based essentially in spin considerations [8] and
group-theoretical argument \textit{a la} Wigner [9]. From the point of view
of $p$-brane theory, however, one has two type of signatures: the background
target spacetime signature, and the worldvolume signature. Thus, the
considerations in references [8] and [9] may be applied to the background
target spacetime signature but not necessarily to the whole scenario of $p$%
-branes.

Through the years there has been some wisdom that such two types of
signatures are connected. For instance, in superstring theory one starts
with $1+1$ signature associated with the worldsheet and supersymmetry and
quantum consistency of the theory requires $1+9$, $5+5$ or $9+1$ background
target spacetime signatures. Similarly, if one starts with $1+2$ worldvolume
signature it is expected that quantum consistency requires $1+10$ background
target spacetime signature. Hull [10] has shown that if one considers closed
time-like dimensions in the worldsheet of string theory or worldvolume $p$%
-branes, then duality implies other signatures for the target spacetime. In
the case of type IIA string theory, Hull suggested that besides the $1+9$
background target spacetime the signatures $0+10,2+8,5+5,4+6$ are also
allowed. On the other hand, a $2+2-$brane signature seems to require $2+10$
background target spacetime signature as it has been shown by several
authors [11]-[14]. Thus, according to these developments our result that $%
1+1 $ and $2+2$ dimensions are exceptional signatures must refer to
worldvolume signature of a $1+1-$brane and $2+2-$brane respectively, rather
than to the background target spacetime signature associated with our World.

The key idea of our proof is based on the observation that a target
spacetime of a $0-$brane may become after first quantization in the
worldvolume of a kind of a $p-$brane [15]-[16]. Thus, $1+1$ and $2+2$ target
spacetime of a $0-$brane can be seen as the $1+1$ and $2+2$ worldvolume
spacetime of a $1+1-$brane and $2+2-$brane respectively. The idea turns out
to be similar to the worldsheets for worldsheets proposed by Green [17], in
which the string worldsheet itself emerges as the embedding space of a two
dimensional string theory, and the $0-$branes condensation suggested by
Townsend [18] in which a $0-$branes condensate leading to a description of a
membrane.

Presumably, $M-$theory should have the property to predict not only the
dimensionality of our World, but also its signature. Further, $M-$theory
should be an exceptional theory and consequently one should expect that
predicts also exceptional $t+s$-branes. In this work we show that the values 
$t=1,s=1$ and $t=2,s=2$ are exceptional and therefore one should expect that 
$1+1$-brane and $2+2$-brane are exceptional $t+s$-branes in agreement with
our results. Moreover, in section 3, using a special kind of $0-$brane
called quatl [15] we show that the signatures $1+1$ and $2+2$ are closely
related. In fact, we show that after quantization a quatl living in $2+2$
target `spacetime' leads to two equivalent branes, namely `$1+1$%
-matrix-brane' and $2+2$-brane. This means that, in principle, one can think
either in $1+1$-matrix-brane or $2+2$-brane as a candidate for the $M-$%
theory. At this respect, it is important to mention that the case of $2+2$%
-brane is precisely what Ketov [19] has suggested as a candidate for the $M-$%
theory.

It is worth mentioning that the relevance of the $2+2$ signature has emerged
in a different context. In mathematics, the signature $2+2$ is known as
Atiyah-Ward signature as a recognition of the work of Atiyah and Ward [20]
(see also [21] and Refs. [22]-[24]) in which they conjecture the possible
classification of lower-dimensional integrable models by means of self-dual
Yang-Mills connection in the $2+2-$signature. Physically, the $2+2-$%
signature emerges in a consistent $N=2$ superstring theory as discussed by
Ooguri and Vafa [25]. These authors proved that $N=2$ strings provide a
consistent quantum theory of self-dual gravity \textit{a la} Plebanski in $%
2+2$ dimensions (see [26]-[27] and Refs. therein). The $N=2$ string concept
has been extended to $N=(2,1)$ heterotic string [28]-[30] and to $N=(2,2)$
open and closed strings [19]. Both of these cases suggested to see $M-$%
theory as a $2+2-$brane embedded in $2+10$ dimensions [11]-[14]. Siegel [31]
has observed that spin-independent coupling and ghost nature of $SO(2,2)$
make N=4 super Yang-Mills in $2+2$ dimensions a topological-like theory.
Finally, it has been emphasized [32]-[33] that Majorana-Weyl spinor exists
in spacetime of $2+2-$signature.

In section 4, we make some comments about the importance of the `$1+1$%
-matrix-brane' or $2+2$-brane for future developments of $M-$theory. In
particular, we argue that the present work may suggest that there should be
an exceptional chirotope connected with the\ $2+2-$signature. (Chirotope
concept is part of the oriented matroid theory [34], which, as it has been
suggested [35]-[39], may provide the underlying mathematical framework for $%
M-$theory, and it may be applied to different scenarios of physics [40] via
the angular momentum concept and bundle matroid theory [41]-[44].) The idea
is essentially suggested because in general, chirotopes and $p$-branes
appear to be connected [39]. From the results of reference [39] it is
straightforward to see that the $t+s-$branes and chirotopes should be also
linked. Therefore, the exceptional feature of the $2+2$-brane should lead
necessarily to an exceptional chirotope.

\bigskip\ 

\textbf{2.- A 0-brane and the signature of the `spacetime'}

\smallskip\ 

Usually, the motion of a $0-$brane is described by the position coordinates $%
x^{\mu }(\tau )$, where $\tau $ is an arbitrary parameter and the index $\mu 
$ goes from $0$ to $3.$ The corresponding lagrangian is

\begin{equation}
L=-m(-\dot{x}^{\mu }\dot{x}^{\nu }\eta _{\mu \nu })^{1/2}.  \tag{1}
\end{equation}%
Here, $m$ is the inertial mass and $\eta _{\mu \nu }=diag(-1,1,1,1)$ is the
Minkowski metric. The signature of the target spacetime determined by $\eta
_{\mu \nu }$ is in this case $1+3.$ It can be proved that the first order
lagrangian

\begin{equation}
L=\dot{x}^{\mu }p_{\mu }-\frac{\lambda }{2}(p^{\mu }p_{\mu }+m^{2}),  \tag{2}
\end{equation}%
where $p_{\mu }$ is the canonical momentum associated with $x^{\mu }$, is
classical equivalent to the lagrangian (1). The advantage of the first order
lagrangian (2) over the lagrangian (1) is that allows the possibility to
consider the massless case

\begin{equation}
L=\dot{x}^{\mu }p_{\mu }-\frac{\lambda }{2}(p^{\mu }p_{\mu }).  \tag{3}
\end{equation}%
Observe that the signature given by $\eta _{\mu \nu }$ is an assumption of
the above problem.

Our main idea is to discuss the above problem from other perspective. First
of all, we shall assume a general signature with $t-$time and $s-$space
coordinates, with $t>0$ and $s>0,$ instead of the usual signature $1+3$.
Secondly, in order to emphasize that the physical system is moving in a more
general target spacetime, we shall change the notation writing $\xi
^{i}(\tau )$ rather than $x^{\mu }(\tau )$ for describing the position of
the system, with $i,j$ running over all values of $t+s$.

Let us rewrite in terms of $\xi ^{i}(\tau )$ the corresponding lagrangian
(3),

\begin{equation}
\mathcal{L}=\dot{\xi}^{i}p_{i}-\frac{\lambda }{2}(p^{i}p_{i}).  \tag{4}
\end{equation}%
Up to total derivative this lagrangian can be written as

\begin{equation}
\mathcal{L}=\frac{1}{2}(\dot{\xi}^{i}p_{i}-\xi ^{i}\dot{p}_{i})-\frac{%
\lambda }{2}(p^{i}p_{i}).  \tag{5}
\end{equation}

Let us now introduce the notation

\begin{equation}
\xi _{1}^{i}\equiv \xi ^{i}  \tag{6}
\end{equation}%
and

\begin{equation}
\xi _{2}^{i}\equiv p^{i}.  \tag{7}
\end{equation}%
Writing $\xi _{1}^{i}$ and $\xi _{2}^{i}$ in the compact form $\xi _{a}^{\mu
}$, with the index $a=1,2,$ we see that (5) becomes%
\begin{equation}
\mathcal{L}=\frac{1}{2}\varepsilon ^{ab}\dot{\xi}_{a}^{i}\xi _{bi}-\frac{%
\lambda }{2}(\xi _{2}^{i}\xi _{2i}),  \tag{8}
\end{equation}%
where $\varepsilon ^{ab}$ is the completely antisymmetric tensor with $%
\varepsilon ^{12}=1.$ It is evident from (8) that the first term is
manifestly $SL(2,R)$ invariant, while the second one is not. Thus, in order
to make the full lagrangian $\mathcal{L}$ manifest $SL(2,R)$ invariant it is
necessary to add new constraints besides the $p^{i}p_{i}=0$ constraint. The
resultant lagrangian is (see [7] and Refs. therein)

\begin{equation}
\mathcal{L}_{0}=\frac{1}{2}\varepsilon ^{ab}\dot{\xi}_{a}^{i}\xi _{bi}-\frac{%
\lambda ^{ab}}{2}(\xi _{a}^{i}\xi _{bi}),  \tag{9}
\end{equation}%
where $\lambda ^{ab}=\lambda ^{ba}$ denotes three different lagrange
multipliers.

The constraints derived from (8) are

\begin{equation}
\xi _{a}^{i}\xi _{bi}=0,  \tag{10}
\end{equation}%
which mean

\begin{equation}
\xi _{1}^{i}\xi _{1i}=0,  \tag{11}
\end{equation}

\begin{equation}
\xi _{1}^{i}\xi _{2i}=0,  \tag{12}
\end{equation}%
and

\begin{equation}
\xi _{2}^{i}\xi _{2i}=0.  \tag{13}
\end{equation}%
According to the notation (6) and (7) this is equivalent to

\begin{equation}
\xi ^{i}\xi ^{j}\eta _{ij}=0,  \tag{14}
\end{equation}

\begin{equation}
\xi ^{i}p^{j}\eta _{ij}=0,  \tag{15}
\end{equation}%
and

\begin{equation}
p^{i}p^{j}\eta _{ij}=0,  \tag{16}
\end{equation}%
respectively (see Ref. [45] for a discussion of these constraints). On the
other hand, applying Noether's procedure to (8) one learns that the angular
momentum

\begin{equation}
L^{ij}=\xi ^{i}p^{j}-\xi ^{j}p^{i}  \tag{17}
\end{equation}%
or

\begin{equation}
L^{ij}=\varepsilon ^{ab}\xi _{a}^{i}\xi _{b}^{j}  \tag{18}
\end{equation}%
is a conserved dynamic variable obeying the Lorentz group algebra.

Our goal is now to determine signatures of $\eta _{ij}$ for which the
formulae (14)-(17) make sense. For this purpose, let us separate from
(14)-(16) one time variable in the form%
\begin{equation}
-(\xi ^{1})^{2}+\xi ^{i^{\prime }}\xi ^{j^{\prime }}\eta _{i^{\prime
}j^{\prime }}=0,  \tag{19}
\end{equation}

\begin{equation}
-\xi ^{1}p^{1}+\xi ^{i^{\prime }}p^{j^{\prime }}\eta _{i^{\prime }j^{\prime
}}=0,  \tag{20}
\end{equation}%
and

\begin{equation}
-(p^{1})+p^{i^{\prime }}p^{j^{\prime }}\eta _{i^{\prime }j^{\prime }}=0, 
\tag{21}
\end{equation}%
where the indices $i^{\prime },j^{\prime }$, etc. run from $2$ to $t+s$. The
formula (20) leads to

\begin{equation}
(\xi ^{1})^{2}(p^{1})^{2}-\xi ^{i^{\prime }}p^{j^{\prime }}\eta _{i^{\prime
}j^{\prime }}\xi ^{k^{\prime }}p^{l^{\prime }}\eta _{k^{\prime }l^{\prime
}}=0.  \tag{22}
\end{equation}%
Using (14) and (16) we find that (22) becomes

\begin{equation}
\xi ^{i^{\prime }}\xi ^{j^{\prime }}\eta _{i^{\prime }j^{\prime
}}p^{k^{\prime }}p^{l^{\prime }}\eta _{k^{\prime }l^{\prime }}-\xi
^{i^{\prime }}p^{j^{\prime }}\eta _{i^{\prime }j^{\prime }}\xi ^{k^{\prime
}}p^{l^{\prime }}\eta _{k^{\prime }l^{\prime }}=0,  \tag{23}
\end{equation}%
which can also be written as

\begin{equation}
(\delta _{i^{\prime }}^{j^{\prime }}\delta _{k^{\prime }}^{l^{\prime
}}-\delta _{i^{\prime }}^{l^{\prime }}\delta _{k^{\prime }}^{j^{\prime
}})\xi ^{i^{\prime }}\xi _{j^{\prime }}p^{k^{\prime }}p_{l^{\prime }}=0. 
\tag{24}
\end{equation}

Let us now introduce the completely antisymmetric symbol

\begin{equation}
\epsilon ^{i_{2}^{\prime }...i_{t+s}^{\prime }}.  \tag{25}
\end{equation}%
This is a rank-$t+s-1$ tensor which values are $+1$ or $-1$ depending on
even or odd permutations of%
\begin{equation}
\epsilon ^{2...t+s},  \tag{26}
\end{equation}%
respectively. Moreover, $\epsilon ^{i_{1}^{\prime }...i_{t+s}^{\prime }}$
takes the value $0,$ unless $i_{2}^{\prime }...i_{t+s}^{\prime }$ are all
different.

The formula (24) can be written in terms of $\epsilon ^{i_{2}^{\prime
}...i_{t+s}^{\prime }}$ in the form

\begin{equation}
\epsilon ^{j^{\prime }l^{\prime }i_{4}^{\prime }...i_{t+s}^{\prime
}}\epsilon _{i^{\prime }k^{\prime }i_{4}^{\prime }...i_{t+s}^{\prime }}\xi
^{i^{\prime }}\xi _{j^{\prime }}p^{k^{\prime }}p_{l^{\prime }}=0,  \tag{27}
\end{equation}%
where we dropped the nonzero factor $\frac{1}{(t+s-2)!}.$ Moreover, the
formula (27) can be rewritten as

\begin{equation}
\epsilon ^{j^{\prime }l^{\prime }i_{4}^{\prime }...i_{t+s}^{\prime
}}\epsilon _{i^{\prime }k^{\prime }i_{4}^{\prime }...i_{t+s}^{\prime
}}L^{i^{\prime }k^{\prime }}L_{j^{\prime }l^{\prime }}=0,  \tag{28}
\end{equation}%
Here, we used (17) and dropped some numerical factors. Observe that $%
L_{i_{4}^{\prime }...i_{t+s}^{\prime }}=\frac{1}{2}\epsilon _{i^{\prime
}k^{\prime }i_{4}^{\prime }...i_{t+s}^{\prime }}L^{i^{\prime }k^{\prime }}$
is the dual of $L^{i^{\prime }k^{\prime }}$.

From (27) we first observe that for the values $t=1$ and zero otherwise the
expression (28) is an euclidean relation and, therefore, implies $%
L_{j^{\prime }l^{\prime }}=0.$ This is true for all values of $s$ except for
the case $s=1,$ which cannot be considered from (28). Indeed, for the case $%
1+1$ we need to go back to (14)-(16) and write,

\begin{equation}
-(\xi ^{1})^{2}+(\xi ^{2})^{2}=0,  \tag{29}
\end{equation}

\begin{equation}
-\xi ^{1}p^{1}+\xi ^{2}p^{2}=0,  \tag{30}
\end{equation}%
and

\begin{equation}
-(p^{1})^{2}+(p^{2})^{2}=0.  \tag{31}
\end{equation}%
One can check, using (29) and (31) that (30) is an identity. Thus (29) and
(31) do not lead to any link between $\xi $ and $p$ and therefore in this
case the angular momentum (17) is well defined. Therefore, we have
discovered that the symmetry $SL(2,R)$ is not consistent with the Lorentz
symmetry for all cases $1+s,$ except $s=1$. This means that for the case of
one time the signature $1+1$ is exceptional.

Another consequence of our analysis is that the minimal dimensionality in
which (28) makes sense is $2+1$, since the other case $1+2$ we have already
discovered that implies $L_{j^{\prime }l^{\prime }}=0.$ Since the formula
(28) is true up to sign we find that the analysis of the cases $t+1$ is
exactly the same that $1+s$ for $t\neq 1$ and $s\neq 1$. Thus, the case $2+1$
also implies $L_{j^{\prime }l^{\prime }}=0$. Hence, this proves that with
two times the minimal case in which the symmetry $SL(2,R)$ is perfectly
consistent with the Lorentz symmetry, corresponds to the $2+2-$signature. In
principle we may continue with this procedure founding that $3+3$ and so on
are consistent possibilities. But, considering that (19)-(21) are only three
constraints we see that there are not enough constraints to eliminate all
additional degrees of freedom in all possible cases with $t\geq 3$ and $%
s\geq 3.$ In fact, one should expect that this will lead to unpleasant
possibilities at the quantum level [7].

Summarizing by imposing the symmetry $SL(2,R)$ and the Lorentz symmetry in
the lagrangian (9) we have shown that the signatures $1+1$ and $2+2$ are
distinguished for any other and therefore they are exceptional.

\bigskip\ 

\textbf{3.- Relation between the signatures} $1+1$ \textbf{and} $2+2$\textbf{%
:} \textbf{the quatl theory}

\smallskip\ 

In this section, we shall show that a particular kind of $0-$brane called
the quatl [15] provides a link between the signatures $1+1$ and $2+2,$ and
also allows the possibility to transfer the target spacetime properties of
the signatures $1+1$ and $2+2$ to a worldvolume properties of $1+1-$brane
and $2+2-$brane respectively. For this purpose let us start addressing the
quatl theory proposed in Ref. [15]. We should mention that some of the
concepts presented here are new and more general that those discussed in
Ref. [15]. In particular, we establish a connection between the quatl theory
and $2+2-$brane theory.

Consider the line element

\begin{equation}
ds^{2}=d\xi ^{\mu }d\xi ^{\nu }\eta _{\mu \nu }.  \tag{32}
\end{equation}%
Here, we shall assume that the indices $\mu ,\nu \in \{1,2,3,4\}$ and that $%
\eta _{\mu \nu }=diag(-1-1,1,1)$ determine the $2+2-$signature$.$ By defining

\begin{equation}
\begin{array}{ccccccc}
\zeta ^{11}\equiv \xi ^{3}, &  & \zeta ^{22}\equiv \xi ^{4}, &  & \zeta
^{12}\equiv \xi ^{1}, &  & \zeta ^{21}\equiv \xi ^{2},%
\end{array}
\tag{33}
\end{equation}%
it is not difficult to show that (23) can also be written as

\begin{equation}
ds^{2}=d\zeta ^{am}d\zeta ^{bn}\eta _{ab}\eta _{mn},  \tag{34}
\end{equation}%
where $a,b,m,n\in \{1,2\}$ and $\eta _{ab}=diag(-1,1)$ and $\eta
_{mn}=diag(-1,1)$. We have that while $\xi ^{\mu }$ are coordinates
associated to a target spacetime of signature $2+2$ the coordinates $\zeta
^{am}$ are associated with a target space time of signature $1+1.$ Thus, the
equivalence between (32) and (34) determines an interesting connection
between the signatures $1+1$ and $2+2$.

Now, consider the transition $\eta _{\mu \nu }\rightarrow G_{\mu \nu }(\xi
^{\alpha })$ in (32), where $G_{\mu \nu }=G_{\nu \mu }$ is a curved
symmetric metric. A natural question is to see which the corresponding
transition in (34) is. One can check that the transitions $\eta
_{ab}\rightarrow \varphi _{1}g_{ab}(\zeta ^{am})$ and $\eta _{mn}\rightarrow
\varphi _{2}\gamma _{mn}(\zeta ^{am})$, where $g_{ab}$ and $\gamma _{mn}$
are two different nonsymmetric metrics and $\varphi _{1}$and $\varphi _{2}$
are two conformal factors, provide a possible identification. between $%
G_{\mu \nu }$ and the two nonsymmetric metrics $g_{ab}(\zeta ^{am})$ and $%
\gamma _{mn}(\zeta ^{am})$. In fact, this can be seen by just observing that
the two metrics $g_{ab}$ and $\gamma _{mn}$ together with the conformal
factors $\varphi _{1}$and $\varphi _{2}$ lead exactly to the same number of
degrees of freedom associated with the symmetric metric $G_{\mu \nu }.$

The next step is to make dynamic the above kinematic relation between the
signatures $1+1$ and $2+2$. For this purpose let us first consider the $%
(t+s) $-brane action

\begin{equation}
S=\frac{1}{2}\tint d\zeta ^{t+s}\sqrt{-G}\left[ G^{AB}\frac{\partial y^{\hat{%
\nu}}}{\partial \zeta ^{A}}\frac{\partial y^{\hat{\sigma}}}{\partial \zeta
^{B}}\gamma _{\hat{\nu}\hat{\sigma}}-((t+s)-2)\right] ,  \tag{35}
\end{equation}%
where we shall assume that $\gamma _{\hat{\nu}\hat{\sigma}}$ is a background
metric associated with some $\sigma -$model in $T+S$ dimensions. From (35)
we obtain the constraint

\begin{equation}
\frac{\partial y^{\hat{\nu}}}{\partial \zeta ^{A}}\frac{\partial y^{\hat{%
\sigma}}}{\partial \zeta ^{B}}\gamma _{\hat{\nu}\hat{\sigma}}-\frac{G_{AB}}{2%
}\left[ G^{CD}\frac{\partial y^{\hat{\nu}}}{\partial \zeta ^{C}}\frac{%
\partial y^{\hat{\sigma}}}{\partial \zeta ^{D}}\gamma _{\hat{\nu}\hat{\sigma}%
}-((t+s)-2)\right] =0.  \tag{36}
\end{equation}%
From (36) we find the expression

\begin{equation}
G^{CD}\frac{\partial y^{\hat{\nu}}}{\partial \zeta ^{C}}\frac{\partial y^{%
\hat{\sigma}}}{\partial \zeta ^{D}}\gamma _{\hat{\nu}\hat{\sigma}}=t+s. 
\tag{37}
\end{equation}

Now, we ask ourselves what the analogue of (36) for a $0-$brane will be.
Presumably, we shall have

\begin{equation}
P_{A}P_{B}-\frac{G_{AB}}{2}\left[ G^{CD}P_{C}P_{D}-\frac{(t+s)-2}{T+S}\right]
=0  \tag{38}
\end{equation}%
and the analogue of (37) will be

\begin{equation}
G^{CD}P_{C}P_{D}=\frac{t+s}{T+S}.  \tag{39}
\end{equation}%
In principle, (38) and (39) can be obtained from (36) and (37) by assuming
that

\begin{equation}
\gamma _{\hat{\mu}\hat{\nu}}=x^{\hat{M}(\hat{A})}\frac{\partial x_{(\hat{A}%
)}^{\hat{N}}}{\partial y^{\hat{\mu}}}x^{\hat{R}(\hat{C})}\frac{\partial x_{(%
\hat{C})}^{\hat{S}}}{\partial y^{\hat{\nu}}}\eta _{\hat{M}\hat{R}}\eta _{%
\hat{N}\hat{S}}.  \tag{40}
\end{equation}%
Here, $x_{(\hat{A})}^{\hat{N}}$ is eigenstate of $P_{C}=-i\frac{\partial }{%
\partial \zeta ^{C}}$ and is an element of $SO(T,S),$ where $T$ and $S$
determine the signature of the flat metric $\eta _{\hat{M}\hat{R}}.$

The corresponding first order lagrangian from which (38) and (39) can be
derived is

\begin{equation}
S=\tint d\tau \left\{ \dot{\zeta}^{A}p_{A}-\sqrt{-G}\left[ G^{AB}P_{A}P_{B}-%
\frac{((t+s)-2)}{T+S}\right] \right\} .  \tag{41}
\end{equation}

Let us now focus in the case of $2+2$ dimensions. We have $t=2$ and $s=2$
and therefore in this case (35) becomes

\begin{equation}
S=\frac{1}{2}\tint d\zeta ^{2+2}\sqrt{-G}\left[ G^{AB}\frac{\partial y^{\hat{%
\nu}}}{\partial \zeta ^{A}}\frac{\partial y^{\hat{\sigma}}}{\partial \zeta
^{B}}\gamma _{\hat{\nu}\hat{\sigma}}-2\right] ,  \tag{42}
\end{equation}%
which corresponds to the bosonic $2+2-$brane action. While the corresponding 
$0-$brane action (41) is reduced to

\begin{equation}
S=\frac{1}{2}\tint d\tau \left\{ \dot{\zeta}^{A}p_{A}-\sqrt{-G}\left[
G^{AB}P_{A}P_{B}-\frac{2}{T+S}\right] \right\} .  \tag{43}
\end{equation}%
For the particular case of $\varphi _{1}=\varphi _{2}=1,$ it is
straightforward to show that this action can be written in terms of $\zeta
^{am}$ and $p_{am}$ as follows:

\begin{equation}
S=\tint d\tau \left[ \dot{\zeta}^{am}p_{am}-\frac{1}{2}\sqrt{-g}\sqrt{%
-\gamma }(g^{ab}\gamma ^{mn}p_{am}p_{bn}-\frac{2}{T+S}\right] ,  \tag{44}
\end{equation}%
where $g_{ab}$ and $\gamma _{mn}$ are the two auxiliary metrics mentioned
above. It turns out that the coordinates $\zeta ^{am}(\tau )$ describe the
position of a quatl [15]. The states $\shortmid x_{(\hat{A})}^{\hat{N}}>$
which allow the connection between the actions (42) and (43) are the so
called ketzal [15]. It is particularly interesting to observe that the
formalism that connects (43) and (44) only works for the signatures $2+2$
and $1+1.$ In fact, assume that $\zeta ^{A}$ is a vector living in a target
spacetime of signature $q+q$, in other words, we consider the same number of
time and space coordinates. Now, consider the possibility to write the
coordinates $\zeta ^{A}$ as square matrix $\zeta ^{am}$ where the maximum
value of the indices $a$ and $m$ is $q.$ It is not difficult to see that
this is possible if $q$ satisfies the formula $q^{2}=q+q$, which has the
unique solution $q=2.$ Thus, a connection of the form (43) and (44) is only
possible for a vector $\zeta ^{A}$ in a target spacetime of $2+2-$signature
and a matrix $\zeta ^{am}$ associated with a $1+1-$signature. It is worth
mentioning that the group theory provides us with an alternative explanation
about the connection between the coordinates $\zeta ^{A}$ and $\zeta ^{am}.$
In fact, the natural global continuous symmetry associated with the
signature $2+2$ is $SO(2,2)$. But due to the isomorphism $SO(2,2)\cong
SL(2,R)^{\prime }\otimes SL(2,R)$ one finds that a $2+2$ dimensional vector $%
\zeta ^{A}$ may be written as $\zeta ^{am},$ where the index $a$ refers to $%
SL(2,R)^{\prime }$ and the index $m$ to $SL(2,R)$ (see Ref. [19]).

From the above observations it becomes evident that (42) may also be written
as

\begin{equation}
S=\frac{1}{2}\tint d\zeta ^{2+2}\sqrt{-g}\sqrt{-\gamma }\left[ g^{ab}\gamma
^{mn}\frac{\partial y^{\hat{\nu}}}{\partial \zeta ^{am}}\frac{\partial y^{%
\hat{\sigma}}}{\partial \zeta ^{an}}\gamma _{\hat{\nu}\hat{\sigma}}-2\right]
,  \tag{45}
\end{equation}%
which presumably it will correspond to the first quantization of the $0-$%
brane called quatl described by the action (44). It seems that action (45)
has not been considered in the literature. For distinguishing from the $2+2-$%
brane, we shall call the physical system described by (45) the `$1+1-$%
matrix-brane'. The action (45) appears interesting because at the quantum
level may allow to follow similar technics to the one used in string theory.
In particular, (45) may allow to determine the critical dimensions of the $%
1+1-$matrix-brane. In principle, one may think on $s+t-$matrix-brane, but
its correspondence with the $t+s-$brane should require more degrees of
freedom for connecting the coordinates $\zeta ^{A}$ and the square matrix $%
\zeta ^{am}$.

\bigskip\ 

\textbf{4.- Some final thoughts about the }$1+1$ and $2+2$ \textbf{signatures%
}

\smallskip\ 

From the above discussion it is evident that in general, one may consider
three different kinds of `spacetimes': the target spacetime ($\mathcal{T}$),
the scenario where a $0-$brane moves, the worldvolume ($\mathcal{W}$)
associated with a $t+s-$brane, and the $T+S$ background target ($\mathcal{BT}
$) spacetime where the $t+s-$brane evolves. We have shown that, in the case
of the signatures $1+1$ and $2+2,$ and only in this case, there is a clear
relation between $\mathcal{T}$ and $\mathcal{W}$. In fact, in this case,
after quantizing the system, the $\mathcal{T}$ of a quatl becomes the $%
\mathcal{W}$ spacetime of $2+2-$brane. This proves that the $2+2-$brane can
be understood as a first quantization of a quatl, a kind of $0-$brane moving
in a spacetime of $1+1-$signature [15]. Traditionally, the $\mathcal{W}$
spacetime of $1+1-$brane determines the $\mathcal{BT-}$spacetime via quantum
consistency and supersymmetrizability. Thus, in the case of $2+2$ one should
expect a connection of the form

\begin{equation}
\mathcal{T}\leftrightarrow \mathcal{W\leftrightarrow BT}.  \tag{46}
\end{equation}%
Supersymmetrizability established that for a $2+2-$brane, the signature of $%
\mathcal{BT}$ should be $2+10$, but a quantum analysis as in the case of
string theory, as far as we know is lacking. Looking the action (45) one
wonders if the signature of $\mathcal{BT}$ for a supersymmetric $2+2-$brane\
should be $2+18$ rather than $2+10.$ The reason for this is that the action
contains two independent metrics in $1+1$ dimensions and one should expect
that each metric leads to $1+9$ dimensions.

The $2+2-$brane in a $\mathcal{BT}$ of $2+10-$spacetime has been proposed as
a candidate of the $M-$theory. But one of the dreams is that $M-$theory is
an exceptional theory. Thus our proof that $2+2$ is an exceptional signature
coincides with this expectation regarding $M-$theory.

Hull [10] has shown that it makes sense to consider compactifications not
only in the spacelike coordinates, but also in the timelike coordinates.
From this point of view one may think in a double dimensional
compactification procedure in order of reducing $2+2$ to $1+1$ dimensions
and simultaneously $2+10$ going to $1+9$. And this is another way to see the
close relation between the signatures $2+2$ and $1+1.$ Another possibility
for a link between the signature $2+2$ and $1+1$ is to apply the
Carlini-Greensite [46] procedure for transforming a spacelike coordinate
into timelike coordinates and vice versa.

Cosmology may also provide the possibility to see the exceptionality of the
signature $2+2.$ In fact, recently it has been shown [47] that
radius-duality in a Kaluza-Klein cosmological models applies to all possible
dimensions $T+S$ in the spacetime of $\mathcal{BT}$ with the only exception
of $2+S.$ One is tempting to speculate that similar technics applied to the
spacetime $\mathcal{W}$ will lead to the exceptionality of the $2+2-$%
signature$.$ This is expected since $2+2-$brane moves naturally $2+10-$%
spacetime which according to duality should be exceptional. Therefore, one
should expect that the cosmological exceptionality of the $2+10$ spacetime
may be translated to the $2+2-$brane new world scenario.

It has been proposed [35]-[40] the so called oriented matroid theory [34] as
the underlaying mathematical structure for $M-$theory. Since $2+2-$brane is
also a candidate for the $M-$theory one should expect a relation between
matroid theory and $2+2-$brane theory. The proof that the signature $2+2$ is
exceptional relied on the combination of both group symmetries $SL(2,R)$ and 
$SO(S,T)$. Specifically one assumes that the $SL(2,R)$-symmetry implies a
nonvanishing angular momentum $L^{i^{\prime }k^{\prime }}$ associated with
the group $SO(S,T)$. But it has been shown [39] that any nonvanishing
angular momentum can be identified with the chirotope concept of the
oriented matroid theory [34]. Hence, it is correct to assure that the $%
SL(2,R)-$symmetry implies a chirotope structure for two times physics. In
particular, according to the result of the present work concerning the
exceptionality of the $2+2-$signature one must have an exceptional chirotope
associated to the $2+2-$brane. In other words, the $SL(2,R)-$symmetry should
imply an exceptional chirotope structure for $M-$theory.

There are a number of possible developments of the present work. By
considering supersymmetry restrictions of the values $T,S,t$ and $s$ in
connection with a superconformal group, Batrachenko, Duff and Lu [48] have
reexamined the $2-$brane case at the end of the De Sitter universe. It may
be interesting to understand the present development from the point of view
of supersymmetry and superconformal group. There has been an attempt to
explain why the $\mathcal{BT-}$spacetime has $1+3-$signature using fractal
technics [49]. It seems interesting to apply fractal technics to the case $%
\mathcal{W-}$spacetime in order to find an alternative explanation of the
exceptionality of the $2+2-$signature. Finally, self-duality seems to be
deeply connected with the $2+2-$signature [50]. Thus it seems also
interesting for further research to investigate self-duality concept at the
level of constraints of the `$1+1-$matrix-brane' system.

\end{document}